\documentstyle[twoside,fleqn,espcrc2,epsfig]{article}

\newcommand{\be}{\begin{equation}}
\newcommand{\ee}{\end{equation}}

\newcommand{\AmS}{{\protect\the\textfont2
  A\kern-.1667em\lower.5ex\hbox{M}\kern-.125emS}}

\title{
\vspace{-3.2cm}
\begin{flushright}
{\normalsize BUHEP-96-18}\\
{\normalsize July 1996}\\
\end{flushright}
\vspace{1.5cm}
Critical behavior and monopole density in U(1) lattice gauge theory
\thanks{Contribution to LATTICE 96, International Symposium on Lattice Field
Theory, St.~Louis, USA. Supported in part under DFG grants Ke 250/7-2 and 
250/12-1 and under DOE grant DE-FG02-91ER40676}}

\author{Werner Kerler\address{Fachbereich Physik, Universit\"at Marburg,\\
D-35032 Marburg, Germany}, Claudio Rebbi\address{Department of Physics,
Boston University,\\Boston, MA 02215, USA}, and
Andreas Weber$^{\mbox{\scriptsize a}}$}

\begin{document}

\begin{abstract}
Our study of the energy distribution has shown that the strength of
the first order transition in the four-dimensional compact U(1)
lattice gauge theory decreases when the coupling $\lambda$ of the
monopole term increases. The disappearance of the energy gap for
sufficiently large values of $\lambda$ indicates that the transition
ultimately becomes of second order. In our present investigation,
based on a finite-size analysis, we show that already at $\lambda=
0.9$ the critical exponent is characteristic of a second-order
transition. Interestingly, this exponent turns out to be definitely
different from that of the Gaussian case. We observe that the monopole
density becomes constant in the second order region. In addition we
find the rather surprising result that the phase transition persists
up to very large values of $\lambda$, where the transition moves to
(large) negative $\beta$.
\end{abstract}

\maketitle

\section{INTRODUCTION}

We investigate the compact U(1) lattice gauge theory in four dimensions
with the Wilson action supplemented by a monopole term \cite{bs85}
\be
S=\beta \sum_{\mu>\nu,x} (1-\cos \Theta_{\mu\nu,x})+
\lambda \sum_{\rho,x} |M_{\rho,x}|
\label{sbl}
\ee
where $M_{\rho,x}=\epsilon_{\rho\sigma\mu\nu}
(\bar{\Theta}_{\mu\nu,x+\sigma}-\bar{\Theta}_{\mu\nu,x}) /4\pi$
and the physical flux $\bar{\Theta}_{\mu\nu,x}\in [-\pi,\pi)$ is related
to the plaquette angle $\Theta_{\mu\nu,x}\in (-4\pi,4\pi)$ by
$\Theta_{\mu\nu,x}=\bar{\Theta}_{\mu\nu,x}+2\pi n_{\mu\nu,x}$ \cite{dt80}.
We use periodic boundary conditions.

Our studies of the energy distribution \cite{krw94,krw95a} have
shown that the strength of the first order transition decreases
with $\lambda$ so that ultimately the transition becomes of second
order.  This observation has enabled us to develop an algorithm by
which simulations have become possible also on big lattices \cite{krw95a}. 
In this work we study the properties of the second order phase transition 
in detail. Because the monopoles play a fundamental role in the dynamics 
of the phase transition, the addition of a monopole term to the action 
appears particularly well suited for exploring critical properties of 
the system appropriate for a continuum limit.

\section{PHASE TRANSITION LINE}

In Ref.~\cite{bs85} Barber and Shrock observed that there is a
shift of the transition point if $\lambda$ is varied.  The effects of
a complete suppression of monopoles have also been studied
\cite{bss85,bmm93}. In our previous investigations we have determined
the location of the phase transition over a wide range of values in
($\beta,\lambda$) space. Now we wish to clarify what happens at very
large $\lambda$. For this purpose we determine the critical points at
values of $\lambda$ substantially larger than previously considered and
allow $\beta$ to take negative values.

In order to keep the computational cost for this study within bearable
limits we have used our topological characterization of the phases
\cite{krw94,krw95}.  It is based on the fact that there is an infinite
network of monopole current lines in the confining phase and no such
network in the Coulomb phase. On finite lattices ``infinite'' is to be
defined in accordance with the boundary conditions \cite{krw96}. For
periodic boundary conditions ``infinite'' is equivalent to
``topologically nontrivial in all directions''.  Since the analysis
of a single configuration (or of a few configurations, because of finite
size effects, when close to the critical point) is already sufficient
to identify the phase, this characterization permits to find the
transition region rather quickly.

It is to be remembered that on finite lattices different order
parameters lead to slightly different critical $\beta$. On an $8^4$
lattice the maximum of the specific heat and our topological order
parameter give values 1.0075(1) and 1.0074(2) for $\lambda=0$, and
0.3870(5) and 0.372(3) for $\lambda=0.9$, respectively. To determine
the location of the maximum of the specific heat for larger $\lambda$
in an efficient way we first determine the critical $\beta$ from the
topological order parameter and then find the maximum of the specific
heat in an easy second step.

In Figure 1 we show our results for the location of the phase
transition $\beta_C$, defined by the maximum of the specific heat, for
values of $\lambda$ ranging up to $1.3$. It can be seen that the phase
transition line continues to negative $\beta$.

Using the topological order parameter we could follow the line of
phase transitions up to still much larger $\lambda$: from $\lambda =
1.4$ where $\beta_C = - 0.52(2)$ to $\lambda$ = 10 where $\beta_C$
is approximately $-1000$. It is to be emphasized that the
characteristic topological properties of the phases have been found to
be fully present throughout the range of the investigation.  Thus we
have the remarkable result that both phases are still present all the
way up to very large $\lambda$.

We observe that finite size effects increase with $\lambda$. This is
indicated by the fact that the transition region becomes broader. The
width of the peak of the specific heat increases by roughly a factor
of 4 from $\lambda = 0$ to $\lambda = 0.9$ and from $\lambda = 0.9$ to
$\lambda = 1.3$. Similarly, from the topological order parameter, we
see an increase by a factor of approximately 6 from $\lambda = 0$ to
$\lambda = 0.9$. A further indication of larger finite size effects is 
that the height of the specific-heat maximum decreases with $\lambda$. The
decrease is roughly by a factor of 15 from $\lambda = 0$ to $\lambda =
0.9$ and from $\lambda = 0.9$ to $\lambda = 1.3$. Thus one also sees
that for very large $\lambda$ a precise determination of the location
of the maximum of the specific heat becomes cumbersome.

\section{MONOPOLE NUMBER DENSITY}

At smaller $\lambda$ we have observed previously \cite{krw94} that the
monopole number density in the Coulomb phase is roughly constant,
while in the confining phase it decreases rapidly towards this
constant with increasing $\lambda$.  Figure 2 gives the monopole
number density along the transition line.  It can be seen that
starting at $\lambda=0.9$ the density becomes constant 
within the errors of the simulation.

 From the behavior exhibited in Figure 2 we conclude that the monopole
number density becomes constant, in the second order region. Indeed
(as pointed out in Sect.~4) the critical behavior characteristic of a
second order transition properly occurs for $\lambda=0.9$ (while for
$\lambda=0.8$ there are still deviations from scaling).
\vspace{10mm}

\section{CRITICAL BEHAVIOR}

In order corroborate our observation that for large $\lambda$ the
phase transition becomes of second order we have investigated the
finite-size scaling behavior of the maximum of the specific heat 
$C_{\mbox{max}}$. It is expected to be
\be
C_{\mbox{max}} \sim L^d
\ee
if the phase transition is of first order and
\be
C_{\mbox{max}} \sim L^{\frac{\alpha}{\nu}}
\ee
if it is of second order, where $\alpha$ is the critical exponent of the
specific heat and $\nu$ the critical exponent of the correlation length.

In Figure 3 we present the simulation results for $C_{\mbox{max}}$  which
we have obtained on lattices with  $L$ = 6, 8, 10, 12 for $\lambda = 0.9$
at the corresponding values of $\beta_C$. The fit to these data gives
\be
{\frac{\alpha}{\nu}} = 0.485(35)
\ee
Clearly this is quite far from 4 and thus the transition not of first order.

\newpage

By the hyperscaling relation $\alpha = 2 - d\,\nu$ we find
\be
\nu = 0.446(5)
\label{nu}
\ee
It is interesting to note that (\ref{nu}) is clearly different
 from the value $\frac{1}{2}$ of the Gaussian case.

The value in Eq.~(\ref{nu}) might be compared with $\nu = 0.36(1)$
obtained recently \cite{hjjln96} in a sphere-like geometry. Further it
should be remembered that values between 0.33 and 0.50 have been
found, in different contexts, in works done in the eighties
\cite{gnc86,l87}.

The critical $\beta$ is expected to behave as
\be
\beta_C(L)=\beta_C(\infty) +a L^{-\frac{1}{\nu}}
\ee
 From this relation, using the value in Eq.~(\ref{nu}) and our data for
$\beta_C(L)$ at $\lambda = 0.9$, we get $\beta_C(\infty)$ = 0.4059(5)
and $a = -1.99(6)$.

We have also performed simulations for $\lambda = 0.8$ on lattices with
$L$ = 6, 8, 10. It turns out that $C_{\mbox{max}}$ does not yet scale in
this case.

As a next step simulations at some value of $\lambda$ larger than 0.9
appear important in order to see whether the value of Eq.~(\ref{nu}) is
universal. In any case our results suggest that a ``line of fixed
points along which a scale-invariant continuum limit will result''
\cite{b81} occurs with the action (1.1) for sufficiently large values of
$\lambda$.

\begin{figure}[tb]
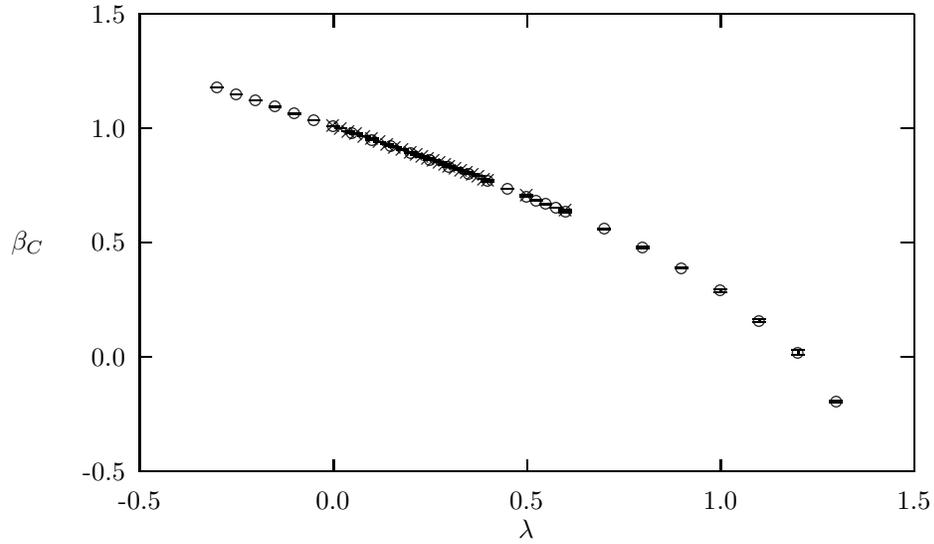

\input fig1.tex
\caption{Location of phase transition $\beta_C$ as function of $\lambda$
for $8^4$ (circles) and $16^4$ (crosses) lattices.}
\end{figure}

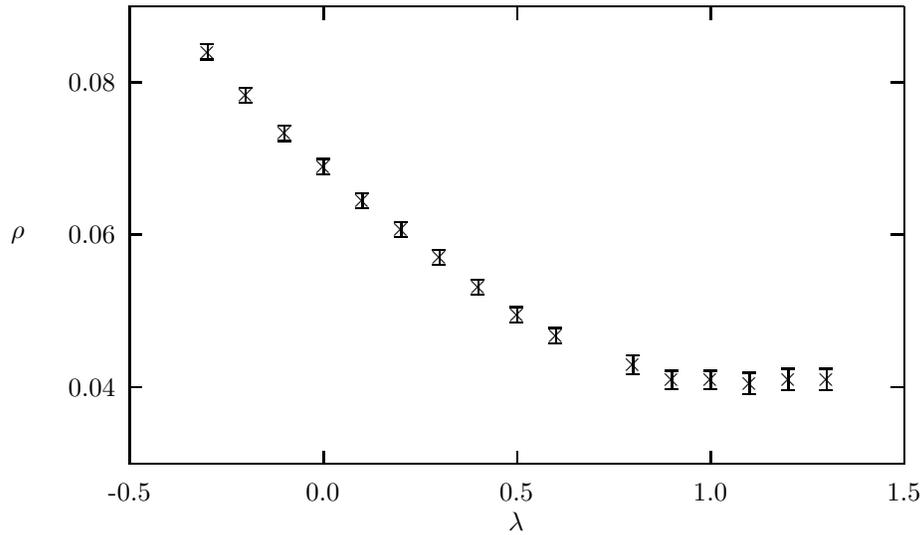
\begin{figure}[tb]
\setlength{\unitlength}{0.240900pt}
\ifx\plotpoint\undefined\newsavebox{\plotpoint}\fi
\begin{picture}(1500,900)(0,0)
\font\gnuplot=cmr10 at 10pt
\gnuplot
\sbox{\plotpoint}{\rule[-0.200pt]{0.400pt}{0.400pt}}%
\put(220.0,712.0){\rule[-0.200pt]{4.818pt}{0.400pt}}
\put(198,712){\makebox(0,0)[r]{0.08}}
\put(1416.0,712.0){\rule[-0.200pt]{4.818pt}{0.400pt}}
\put(220.0,473.0){\rule[-0.200pt]{4.818pt}{0.400pt}}
\put(198,473){\makebox(0,0)[r]{0.06}}
\put(1416.0,473.0){\rule[-0.200pt]{4.818pt}{0.400pt}}
\put(220.0,233.0){\rule[-0.200pt]{4.818pt}{0.400pt}}
\put(198,233){\makebox(0,0)[r]{0.04}}
\put(1416.0,233.0){\rule[-0.200pt]{4.818pt}{0.400pt}}
\put(220.0,113.0){\rule[-0.200pt]{0.400pt}{4.818pt}}
\put(220,68){\makebox(0,0){-0.5}}
\put(220.0,812.0){\rule[-0.200pt]{0.400pt}{4.818pt}}
\put(524.0,113.0){\rule[-0.200pt]{0.400pt}{4.818pt}}
\put(524,68){\makebox(0,0){0.0}}
\put(524.0,812.0){\rule[-0.200pt]{0.400pt}{4.818pt}}
\put(828.0,113.0){\rule[-0.200pt]{0.400pt}{4.818pt}}
\put(828,68){\makebox(0,0){0.5}}
\put(828.0,812.0){\rule[-0.200pt]{0.400pt}{4.818pt}}
\put(1132.0,113.0){\rule[-0.200pt]{0.400pt}{4.818pt}}
\put(1132,68){\makebox(0,0){1.0}}
\put(1132.0,812.0){\rule[-0.200pt]{0.400pt}{4.818pt}}
\put(1436.0,113.0){\rule[-0.200pt]{0.400pt}{4.818pt}}
\put(1436,68){\makebox(0,0){1.5}}
\put(1436.0,812.0){\rule[-0.200pt]{0.400pt}{4.818pt}}
\put(220.0,113.0){\rule[-0.200pt]{292.934pt}{0.400pt}}
\put(1436.0,113.0){\rule[-0.200pt]{0.400pt}{173.207pt}}
\put(220.0,832.0){\rule[-0.200pt]{292.934pt}{0.400pt}}
\put(45,472){\makebox(0,0){$\rho$}}
\put(828,23){\makebox(0,0){$\lambda$}}
\put(828,877){\makebox(0,0){ }}
\put(950,12935){\makebox(0,0)[l]{ }}
\put(220.0,113.0){\rule[-0.200pt]{0.400pt}{173.207pt}}
\put(342,760){\makebox(0,0){$\times$}}
\put(402,692){\makebox(0,0){$\times$}}
\put(463,632){\makebox(0,0){$\times$}}
\put(524,580){\makebox(0,0){$\times$}}
\put(585,526){\makebox(0,0){$\times$}}
\put(646,481){\makebox(0,0){$\times$}}
\put(706,437){\makebox(0,0){$\times$}}
\put(767,390){\makebox(0,0){$\times$}}
\put(828,347){\makebox(0,0){$\times$}}
\put(889,314){\makebox(0,0){$\times$}}
\put(1010,269){\makebox(0,0){$\times$}}
\put(1071,245){\makebox(0,0){$\times$}}
\put(1132,245){\makebox(0,0){$\times$}}
\put(1193,239){\makebox(0,0){$\times$}}
\put(1254,245){\makebox(0,0){$\times$}}
\put(1314,245){\makebox(0,0){$\times$}}
\put(342.0,748.0){\rule[-0.200pt]{0.400pt}{5.782pt}}
\put(332.0,748.0){\rule[-0.200pt]{4.818pt}{0.400pt}}
\put(332.0,772.0){\rule[-0.200pt]{4.818pt}{0.400pt}}
\put(402.0,680.0){\rule[-0.200pt]{0.400pt}{5.782pt}}
\put(392.0,680.0){\rule[-0.200pt]{4.818pt}{0.400pt}}
\put(392.0,704.0){\rule[-0.200pt]{4.818pt}{0.400pt}}
\put(463.0,620.0){\rule[-0.200pt]{0.400pt}{5.782pt}}
\put(453.0,620.0){\rule[-0.200pt]{4.818pt}{0.400pt}}
\put(453.0,644.0){\rule[-0.200pt]{4.818pt}{0.400pt}}
\put(524.0,568.0){\rule[-0.200pt]{0.400pt}{5.782pt}}
\put(514.0,568.0){\rule[-0.200pt]{4.818pt}{0.400pt}}
\put(514.0,592.0){\rule[-0.200pt]{4.818pt}{0.400pt}}
\put(585.0,514.0){\rule[-0.200pt]{0.400pt}{5.782pt}}
\put(575.0,514.0){\rule[-0.200pt]{4.818pt}{0.400pt}}
\put(575.0,538.0){\rule[-0.200pt]{4.818pt}{0.400pt}}
\put(646.0,469.0){\rule[-0.200pt]{0.400pt}{5.782pt}}
\put(636.0,469.0){\rule[-0.200pt]{4.818pt}{0.400pt}}
\put(636.0,493.0){\rule[-0.200pt]{4.818pt}{0.400pt}}
\put(706.0,425.0){\rule[-0.200pt]{0.400pt}{5.782pt}}
\put(696.0,425.0){\rule[-0.200pt]{4.818pt}{0.400pt}}
\put(696.0,449.0){\rule[-0.200pt]{4.818pt}{0.400pt}}
\put(767.0,378.0){\rule[-0.200pt]{0.400pt}{5.782pt}}
\put(757.0,378.0){\rule[-0.200pt]{4.818pt}{0.400pt}}
\put(757.0,402.0){\rule[-0.200pt]{4.818pt}{0.400pt}}
\put(828.0,335.0){\rule[-0.200pt]{0.400pt}{5.782pt}}
\put(818.0,335.0){\rule[-0.200pt]{4.818pt}{0.400pt}}
\put(818.0,359.0){\rule[-0.200pt]{4.818pt}{0.400pt}}
\put(889.0,302.0){\rule[-0.200pt]{0.400pt}{5.782pt}}
\put(879.0,302.0){\rule[-0.200pt]{4.818pt}{0.400pt}}
\put(879.0,326.0){\rule[-0.200pt]{4.818pt}{0.400pt}}
\put(1010.0,254.0){\rule[-0.200pt]{0.400pt}{6.986pt}}
\put(1000.0,254.0){\rule[-0.200pt]{4.818pt}{0.400pt}}
\put(1000.0,283.0){\rule[-0.200pt]{4.818pt}{0.400pt}}
\put(1071.0,230.0){\rule[-0.200pt]{0.400pt}{6.986pt}}
\put(1061.0,230.0){\rule[-0.200pt]{4.818pt}{0.400pt}}
\put(1061.0,259.0){\rule[-0.200pt]{4.818pt}{0.400pt}}
\put(1132.0,230.0){\rule[-0.200pt]{0.400pt}{6.986pt}}
\put(1122.0,230.0){\rule[-0.200pt]{4.818pt}{0.400pt}}
\put(1122.0,259.0){\rule[-0.200pt]{4.818pt}{0.400pt}}
\put(1193.0,222.0){\rule[-0.200pt]{0.400pt}{8.191pt}}
\put(1183.0,222.0){\rule[-0.200pt]{4.818pt}{0.400pt}}
\put(1183.0,256.0){\rule[-0.200pt]{4.818pt}{0.400pt}}
\put(1254.0,228.0){\rule[-0.200pt]{0.400pt}{8.191pt}}
\put(1244.0,228.0){\rule[-0.200pt]{4.818pt}{0.400pt}}
\put(1244.0,262.0){\rule[-0.200pt]{4.818pt}{0.400pt}}
\put(1314.0,228.0){\rule[-0.200pt]{0.400pt}{8.191pt}}
\put(1304.0,228.0){\rule[-0.200pt]{4.818pt}{0.400pt}}
\put(1304.0,262.0){\rule[-0.200pt]{4.818pt}{0.400pt}}
\end{picture}
\caption{Monopole density as function of $\lambda$.}
\end{figure}
\clearpage

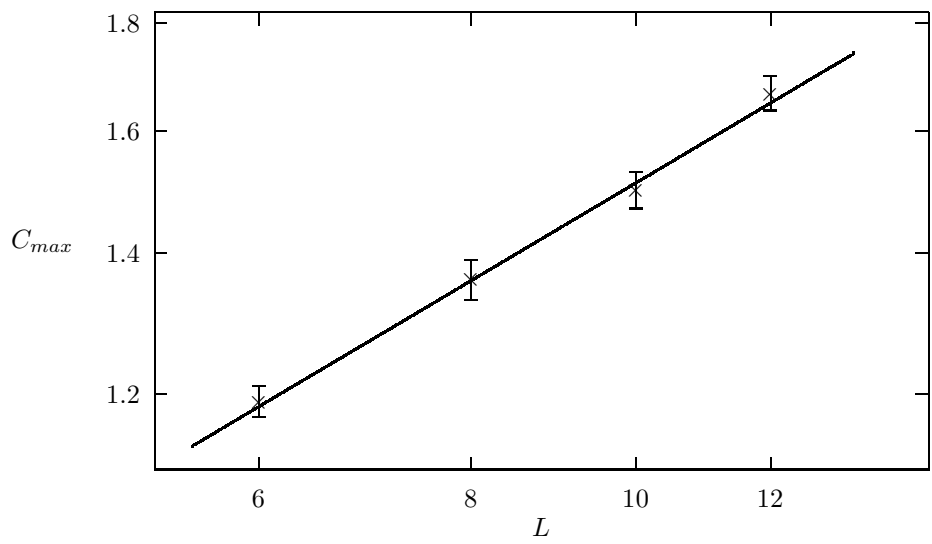
\begin{figure}[tb]
\setlength{\unitlength}{0.240900pt}
\ifx\plotpoint\undefined\newsavebox{\plotpoint}\fi
\begin{picture}(1500,900)(0,0)
\font\gnuplot=cmr10 at 10pt
\gnuplot
\sbox{\plotpoint}{\rule[-0.200pt]{0.400pt}{0.400pt}}%
\put(220.0,231.0){\rule[-0.200pt]{4.818pt}{0.400pt}}
\put(198,231){\makebox(0,0)[r]{1.2}}
\put(1416.0,231.0){\rule[-0.200pt]{4.818pt}{0.400pt}}
\put(220.0,453.0){\rule[-0.200pt]{4.818pt}{0.400pt}}
\put(198,453){\makebox(0,0)[r]{1.4}}
\put(1416.0,453.0){\rule[-0.200pt]{4.818pt}{0.400pt}}
\put(220.0,645.0){\rule[-0.200pt]{4.818pt}{0.400pt}}
\put(198,645){\makebox(0,0)[r]{1.6}}
\put(1416.0,645.0){\rule[-0.200pt]{4.818pt}{0.400pt}}
\put(220.0,814.0){\rule[-0.200pt]{4.818pt}{0.400pt}}
\put(198,814){\makebox(0,0)[r]{1.8}}
\put(1416.0,814.0){\rule[-0.200pt]{4.818pt}{0.400pt}}
\put(384.0,113.0){\rule[-0.200pt]{0.400pt}{4.818pt}}
\put(384,68){\makebox(0,0){6}}
\put(384.0,812.0){\rule[-0.200pt]{0.400pt}{4.818pt}}
\put(717.0,113.0){\rule[-0.200pt]{0.400pt}{4.818pt}}
\put(717,68){\makebox(0,0){8}}
\put(717.0,812.0){\rule[-0.200pt]{0.400pt}{4.818pt}}
\put(976.0,113.0){\rule[-0.200pt]{0.400pt}{4.818pt}}
\put(976,68){\makebox(0,0){10}}
\put(976.0,812.0){\rule[-0.200pt]{0.400pt}{4.818pt}}
\put(1187.0,113.0){\rule[-0.200pt]{0.400pt}{4.818pt}}
\put(1187,68){\makebox(0,0){12}}
\put(1187.0,812.0){\rule[-0.200pt]{0.400pt}{4.818pt}}
\put(220.0,113.0){\rule[-0.200pt]{292.934pt}{0.400pt}}
\put(1436.0,113.0){\rule[-0.200pt]{0.400pt}{173.207pt}}
\put(220.0,832.0){\rule[-0.200pt]{292.934pt}{0.400pt}}
\put(45,472){\makebox(0,0){$C_{max}$}}
\put(828,23){\makebox(0,0){$L$}}
\put(828,877){\makebox(0,0){ }}
\put(220.0,113.0){\rule[-0.200pt]{0.400pt}{173.207pt}}
\put(384,219){\makebox(0,0){$\times$}}
\put(717,412){\makebox(0,0){$\times$}}
\put(976,552){\makebox(0,0){$\times$}}
\put(1187,703){\makebox(0,0){$\times$}}
\put(384.0,195.0){\rule[-0.200pt]{0.400pt}{11.804pt}}
\put(374.0,244.0){\rule[-0.200pt]{4.818pt}{0.400pt}}
\put(374.0,195.0){\rule[-0.200pt]{4.818pt}{0.400pt}}
\put(717.0,379.0){\rule[-0.200pt]{0.400pt}{15.177pt}}
\put(707.0,442.0){\rule[-0.200pt]{4.818pt}{0.400pt}}
\put(707.0,379.0){\rule[-0.200pt]{4.818pt}{0.400pt}}
\put(976.0,523.0){\rule[-0.200pt]{0.400pt}{13.731pt}}
\put(966.0,580.0){\rule[-0.200pt]{4.818pt}{0.400pt}}
\put(966.0,523.0){\rule[-0.200pt]{4.818pt}{0.400pt}}
\put(1187.0,677.0){\rule[-0.200pt]{0.400pt}{13.009pt}}
\put(1177.0,731.0){\rule[-0.200pt]{4.818pt}{0.400pt}}
\put(1177.0,677.0){\rule[-0.200pt]{4.818pt}{0.400pt}}
\sbox{\plotpoint}{\rule[-0.400pt]{0.800pt}{0.800pt}}%
\put(278,149){\usebox{\plotpoint}}
\multiput(278.00,150.41)(0.842,0.500){1231}{\rule{1.547pt}{0.120pt}}
\multiput(278.00,147.34)(1038.790,619.000){2}{\rule{0.773pt}{0.800pt}}
\end{picture}
\caption{$C_{\mbox{max}}$ versus $L$ for $\lambda = 0.9$ at $\beta_C$.}
\end{figure}

\end{document}